# Large-Scale Screening of COVID-19 from Community Acquired Pneumonia using Infection Size-Aware Classification

Feng Shi[†], Liming Xia[†], Fei Shan[†], Dijia Wu, Ying Wei, Huan Yuan, Huiting Jiang, Yaozong Gao, He Sui, Dinggang Shen[*]

*Abstract*— The worldwide spread of coronavirus disease (COVID-19) has become a threatening risk for global public health. It is of great importance to rapidly and accurately screen patients with COVID-19 from community acquired pneumonia (CAP). In this study, a total of 1658 patients with COVID-19 and 1027 patients of CAP underwent thin-section CT. All images were preprocessed to obtain the segmentations of both infections and lung fields, which were used to extract location-specific features. An infection Size Aware Random Forest method (iSARF) was proposed, in which subjects were automated categorized into groups with different ranges of infected lesion sizes, followed by random forests in each group for classification. Experimental results show that the proposed method yielded sensitivity of 0.907, specificity of 0.833, and accuracy of 0.879 under five-fold cross-validation. Large performance margins against comparison methods were achieved especially for the cases with infection size in the medium range, from 0.01% to 10%. The further inclusion of Radiomics features show slightly improvement. It is anticipated that our proposed framework could assist clinical decision making.

## I. Introduction

As known as COVID-19, coronavirus disease outbreaks in Wuhan, China since Dec. 2019, and rapidly spreads nationwide and globally[1, 2]. As a form of pneumonia, the infection causes inflammation of air sacs in one or both lungs and fills with fluid or pus, making the patient difficult to breathe. According to recent report, its lethal rate reaches 2%[3], second only to Severe Acute Respiratory Syndrome (SARS) of 10% and Middle East Respiratory Syndrome (MERS) of 36%[4, 5]. Also, it could transmit from person to person with a relatively high basic reproduction number ($R_0$) of 2.2, and has no efficient treatments and control strategies till now[6]. Therefore, screening of COVID-19 out of community acquired pneumonia (CAP) is important for patient triage, treatment protocol design, and follow-up evaluation.

Currently, a patient, showing respiratory symptoms, fever, cough, dyspnea, or pneumonia, would be referred to the test of real-time Polymerase Chain Reaction (RT-PCR) for the final diagnosis of COVID-19[7]. However, recent studies show that RT-PCR has relatively low detection rate at around 30-60%[8, 9], and repeated tests are generally needed. Therefore, negative results of RT-PCR could not rule out the possibility of infection. Chest CT is a routine diagnostic tool for pneumonia, and is found very useful in detecting typical radiographic features of COVID-19, especially with thin slices[8, 10]. Basically, these features include bilateral and peripheral ground-glass and also consolidative pulmonary opacities. According to the time course of disease development, patients could be divided into mild, moderate, severe and critically ill stages[11]. In the mild stage, almost no pneumonia could be seen from CT images. From common to critically ill stages, one could observe ground-glass opacity (GGO), increased crazy-paving pattern, and consolidation[8]. If the symptom of patient improves, gradual resolution of consolidation could be seen in CT images. However, it remains a great challenge in screening of COVID-19 from typical vital pneumonia in CT images. First, same radiographic features presenting similar appearances to other types of pneumonia, makes it difficult to differentiate from each other. Second, CT images, especially for thin slice acquisition, contain hundreds of slices for manual checking, which is time-consuming and may produce false negative results especially at the mild disease stage. Accordingly, an automated image analysis method for classifying COVID-19 from CAP is highly desired.

Although there are a number of researches summarizing typical CT radiographic signs in COVID-19 patients for guiding clinical practice, the study on automated machine learning assisted disease screening is still limited. For example, Wang et al. proposed a deep learning method to classify the patches of infected lesions into COVID-19 or typical viral pneumonia of totally 99 subjects, where manually labeled infections were required [12]. Xu *et al.* studied the early screening of COVID-19 from Influenza-A viral pneumonia and healthy cases with totally 618 CT samples[13], where lesion patches were extracted and combined with relative distance-from-edge features for diagnosis. Li et al. proposed to a 2D model to extract features from 2D slices and combined to provide diagnosis of 468 COVID-19, 1511 CAP, and 1303 non-pneumonia patients.

[†] F. Shi, L. Xia, and F. Shan contributed equally to this work.
[*] Corresponding author. Email: dinggang.shen@gmail.com
F. Shi, D. Wu, Y. Wei, H. Yuan, H. Jiang, Y. Gao, and D. Shen are with the Department of Research and Development, Shanghai United Imaging Intelligence Co., Ltd., Shanghai, China.
L. Xia is with Department of Radiology, Tongji Hospital, Tongji Medical College, Huazhong University of Science and Technology, Wuhan, Hubei, China.
F. Shan is with Department of Radiology, Shanghai Public Health Clinical Center, Fudan University, Shanghai, China.
H. Sui is with Department of Radiology, China-Japan Union Hospital of Jilin University, Changchun, China.

Another difficulty in differentiating COVID-19 patients from CAP patients is the vast difference of distribution of disease stages in the available data. For COVID-19, the stages from mild condition to critically ill almost evenly distributed in the population of COVID-19 patients. However, for CAP, most cases remain in mild to intermediate degree of pneumonia as observed in CT images. The main reason could be that there are already effective treatments available for CAP while not for COVID-19 yet. This leads to inconsistent distribution of disease severity in the acquired dataset where small infections found in CT images are most likely belonging to CAP, while wide spread infections found in CT images are for COVID-19. A general classification algorithm may take advantage of this situation and report an overall high performance result, but may have low sensitivity and specificity in the middle range where the infection size is not too small or too large, which helps little for doctors in clinical practice. This is mainly because the effect of the size of infection has not been evaluated and properly considered in the classification algorithm.

In this paper, we propose a machine learning approach for screening COVID-19 out of CAP. Our contributions are 3-folds. 1) We presented an infection size-aware method, and broke down the evaluation of COVID-19 classification task into multiple infection size ranges, respectively. 2) We proposed a location-specific feature extraction process, where COVID-19 features were collected and tailored according to current understanding of radiographic appearance. 3) The proposed method was throughout evaluated on a large-scale dataset from multiple centers, covering patients with age ranging from 12 to 98 years. Details are provided in the following sections.

## II. MATERIALS AND METHODS

### A. Participants and image acquisition

CT images of a total of 2685 participants were retrospectively collected. In this dataset, 1658 cases were the confirmed COVID-19 cases diagnosed by positive nucleic acid testing with conformation by national CDC. The other 1027 cases were CAP patients. Three hospitals were involved, including Tongji Hospital of Huazhong University of Science and Technology, Shanghai Public Health Clinical Center of Fudan University, and China-Japan Union Hospital of Jilin University.

All patients underwent chest CT scans with thin section. Specifically, CT scanners include uCT 780 from UIH, Optima CT520, Discovery CT750, LightSpeed 16 from GE, Aquilion ONE from Toshiba, SOMATOM Force from Siemens, and SCENARIA from Hitachi. CT protocol includes: 120 kV, reconstructed CT thickness ranges from 0.625 to 2mm, with breath hold at full inspiration. All images were de-identified before sending for analysis. The study was approved by the Institutional Review Board of participating institutes. Written informed consent was waived due to the retrospective nature of the study. A mediastinal window (with window width 350 HU and window level 40 HU) and lung window (with window width 1200 HU and window level-600 HU) were used for reading.

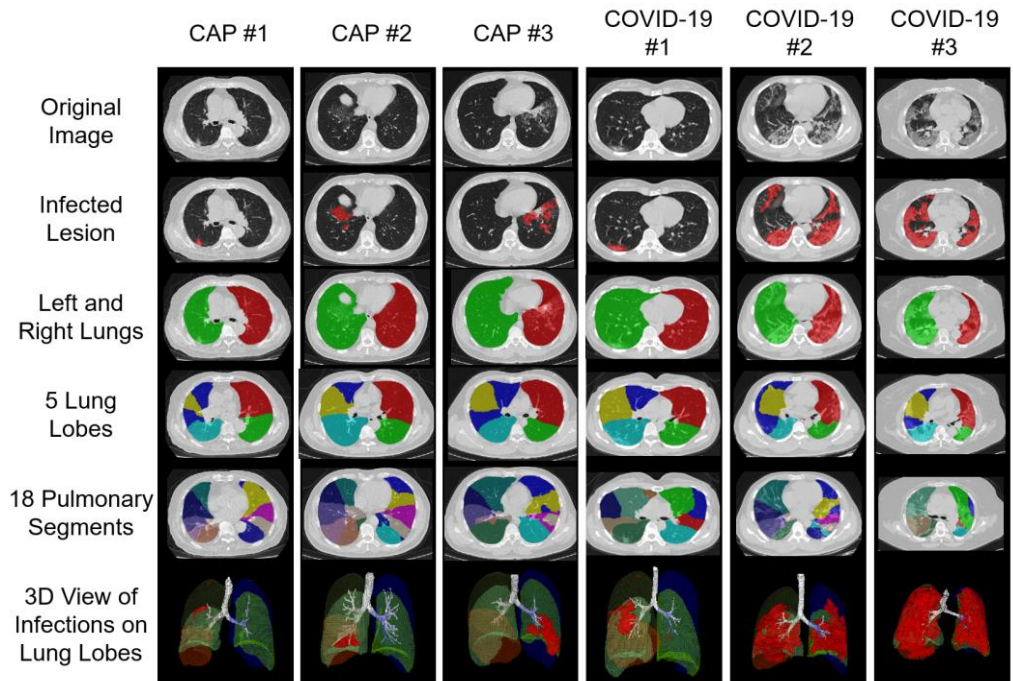

Fig. 1. Illustration of lung images, preprocessed results of infected lesions and lung fields on 3 CAP (left 3 columns) and 3 COVID-19 patients (right 3 columns), respectively.

## B. Preprocess of Lung Infections and Fields

We propose to utilize the disease characteristics, i.e., infection locations and spreading patterns, to extract handcrafted features. To do that, we automatically segmented infected lung regions and lung fields bilaterally. The infected lung regions were mainly related to manifestations of pneumonia, such as mosaic sign, GGO, lesion-related signs (air bronchogram), and interlobular septal thickening. The resulting lung fields include left and right lungs, five lung lobes, and eighteen pulmonary segments.

The segmentation process was done through our in-house research portal software[14]. Specifically, a deep learning based network called VB-Net was employed for image segmentation. The VB-Net is a modified network that combines V-Net with bottleneck layers to reduce and combine feature map channels[15]. The network includes a contracting path to extract global image features and an expansive path to integrate fine-grained image features. A bottleneck structure is integrated in the network to reduce the number of feature map channels and thus speed up the spatial convolution. Given proper annotations as training data, the V-Net is capable of segmenting infected lesions as well as lung fields. This software has been evaluated in segmentation of infections and lung fields and achieved Dice similarity coefficient of 92% between automated and manual infection segmentations. In this work, we directly use this software for segmentation of infected lesions and lung fields.

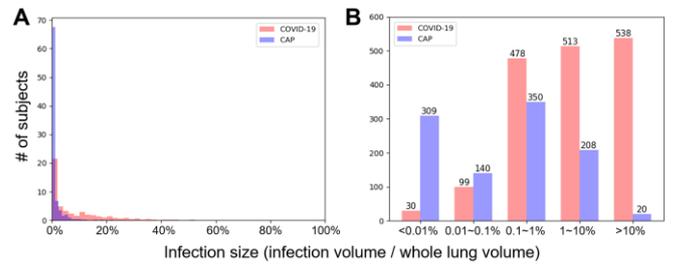

Fig. 2. Illustration of size distribution in patients. (A) shows a highly skewed distribution for the number of subjects as a function of infection size. (B) separates the dataset into 5 groups with exponential size ranges.

An example of images and preprocessed results is shown in Fig. 1. A total of 3 CAP and 3 COVID-19 patients are included in 6 columns, respectively. Segmentations of infected lesions and lung fields are also shown.

## C. Infection Size Distribution and Proposed Size-Aware Method

We refer the infection size as the volume of infected regions against the volume of whole segmented lung. Fig. 2A shows the distribution of infection size in patients. As the relationship between size and disease progress is not linear, we also tried to separate patients into 4 groups, based on exponential intervals. Results show that Fig. 2B can better separate the patients. As can observe that, COVID-19 patients in our dataset tend to have larger infections. For example, they account for 96% (538 vs. 20) in the size group of >10%.

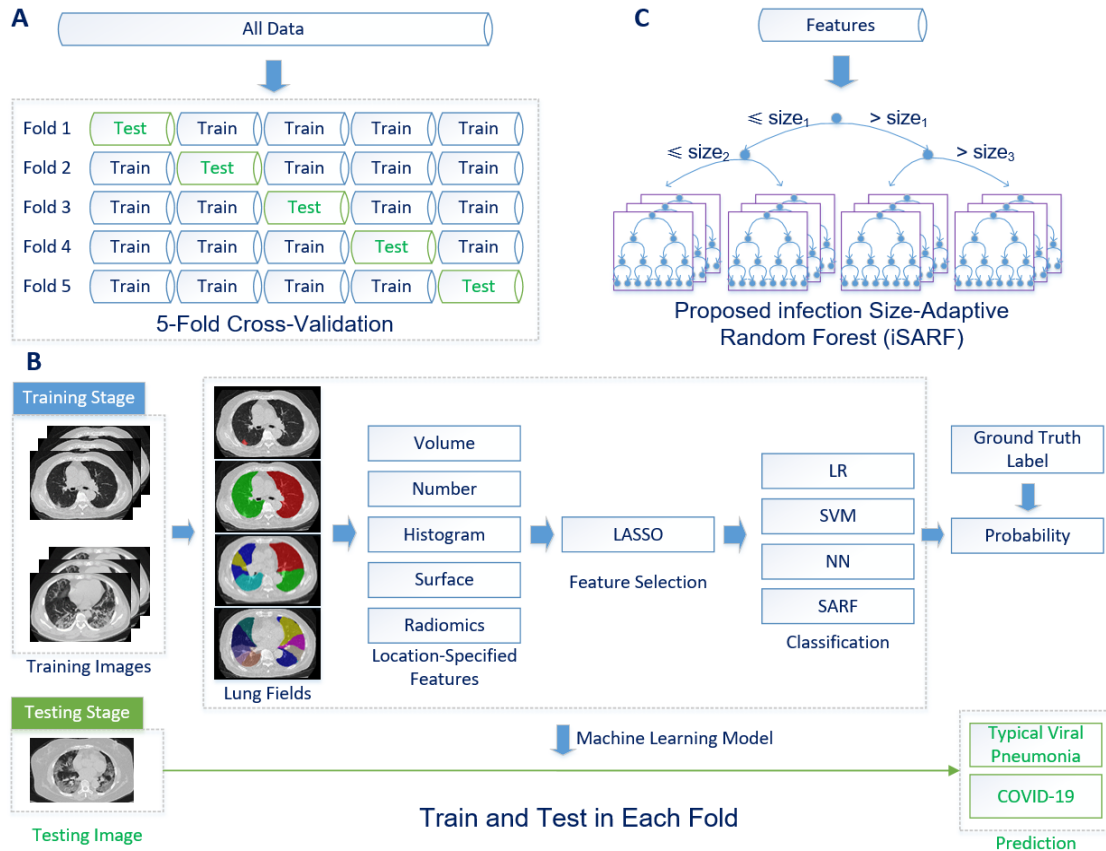

Fig. 3. Flow chart of (A) the 5-fold cross validation, (B) train and test process in each fold, and (C) the proposed size-aware method.

In contrast, CAP patients occupy 91% (309 vs. 30) in the size group of <0.01%.

As above mentioned, such polarized patient distribution is not ideal for conventional classification models. In these models, size would be chosen as a major feature as it could easily separate both groups, although this leads to a low performance for the classification of middle-size groups and does not reflect the real radiographic appearance differences between two types of pneumonia. In this work, we proposed a new classification strategy with its flow chart shown in Fig. 3. In particular, all data were randomly partitioned to 5 equally-sized subsamples. In each fold, 4 subsamples were used for training data to construct the classification model, and the rest one subsample was used as testing data. This process was repeated until all subsamples were tested. Fig. 3B shows the schematic diagram where, in each fold, all images were preprocessed, and machine learning models were trained including feature extraction, selection, and disease classification. In the testing stage, a new image would undergo the preprocessing steps, and then the model predicted its probability of being COVID-19 or CAP.

Herein, we proposed an infection size-aware random forest (iSARF) method (Fig. 3C). From all features, the infection size was used as the only feature in a 3-level random forest where a decision tree was formed and used to separate the data into 4 size groups. Then a group of random forests were constructed for each size group in the training process. In the testing stage, the testing data would be sent to a proper size group through the decision tree and then classified by the following (respective) random forests for the final diagnosis. Details of feature extraction and classification are provided in the following subsections.

*D. Extraction of Location-Specific Features*

In this study, we proposed a series of handcrafted features to be automatically extracted in CT images from infections and lung fields. These features are composed of 4 categories, including the volume, infected lesion number, histogram distribution, and surface area. The detailed feature distribution framework is shown in Fig. 4.

**Volume features**. We extracted the total volume of infected region, and calculated the percentage of the infected region of the whole lung. By using the lung field masks, we further extract the volume and percentage, respectively, in each lobe and pulmonary segment. As there is evidence that COVID-19 more likely occur in both lungs, we calculate the infected lesion difference as well as the percentage difference between left and right lungs.

**Infected lesion number**. Another image difference between COVID-19 and CAP is that most of COVID-19 infections encompassed bilateral lungs with multifocal involvement[8, 16], and COVID-19 generally also has concentrated infected lesions while CAP shows small in volume and patchy in distribution[17]. Therefore, we calculate the characteristics of the total number of infected regions in the bilateral lungs, lung lobes, and pulmonary segments, respectively.

**Histogram distribution**. The manifestations of COVID-19 had its own characteristics, which are different from other types of pneumonia, such as Influenza-A viral pneumonia[13]. The predominant chest CT findings show that bilateral and peripheral GGO and consolidation were a radiologic hallmark of COVID-19[6, 18, 19]. GGO is a pattern of hazy increased lung opacity with preservation of bronchial and vascular margins, whereas consolidation is characterized by a homogeneous increase in lung parenchymal attenuation that obscures the margins of vessels and airway walls on CT image[17]. In order to extract CT intensity distribution of the infected region, we calculated the histogram features of the infected region. According to the window width of the lung window of 1500 and the window level of -600, the intensity value range of the lung region image obtained by inference tends to be between -1350 and 150. According to the preset intensity value interval, we divided the interval into 30 equal bins, and counted the frequency of intensity level in the infected region at each bin to obtain the frequency distribution histogram features.

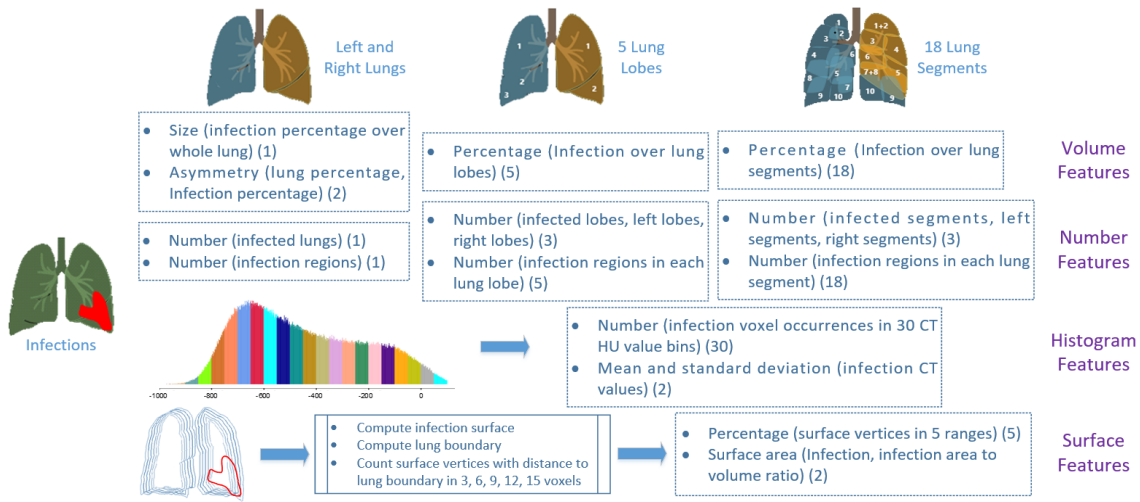

Fig. 4. Illustration of the extraction of handcrafted image features. There are totally 96 features, including 26 volume features, 31 number features, 32 histogram features, and 7 surface features.

**Surface area**. Compared with CAP, it has been found that COVID-19 had a predominate distribution in the posterior and peripheral lung[20], and the abnormalities of lung parenchyma eventually spread to the central area and bilateral upper lobes[8]. Therefore, we constructed the infection surface as well as the lung boundary surface. We further calculated the distance of each infection surface vertex to the nearest lung boundary surface, and categorized them into 5 ranges, as 3, 6, 9, 12, and 15 voxels (voxel spacing is 1.5mm). For features, the number of infection surface vertices within each range of distances to the lung wall were calculated. Furthermore, the percentage of infection vertex number against the number of whole infection surface vertices in each range were also obtained.

*E. Feature Selection and Prediction*

After generating the features, we apply machine-learning methods to select proper features, and predict COVID-19 patients from CAP patients. The feature selection and prediction process of each fold are detailed as follows.

In the training stage, we employed least absolute shrinkage and selection operator (LASSO) to explore the optimal subset of clinico-radiological features for the classification, due to its ability to provide variable importance and interpretability. After that, the selected features were fed into LR, SVM, NN, and the proposed method, respectively, to determine their hyperparameters for disease diagnosis. For LR, we used with default parameters of penalty as L2 (the norm used in the penalization) and regularization C as 1. For SVM, we used radial basis function (RBF) kernel with parameters of regularization C and gamma being determined through an inner 5-fold grid search. For NN, we used MLP Classifier with one hidden layer of 100 nodes and with max iterations of 500. For the proposed method, 100 trees of random forest were used for each size group, maximum depth of tree is 10, and Gini impurity is employed to measure the quality of a split. These trained models were then applied to new test images to predict their probability of being COVID-19 against CAP in the testing stage.

To evaluate the performance, receiver operating characteristic (ROC) analysis was performed. Sensitivity evaluates the ratio of correctly identified positive cases versus all positive cases. Specificity measures the ratio of negatives correctly found in all true negatives. Area under the curve (AUC) demonstrates the ability of classifier in consideration of both sensitivity and specificity. The most weighted features were listed. We further investigated the metrics of results with respect to the infected lesion size.

### III. RESULTS

The dataset includes 1658 patients with COVID-19 and 1027 patients with CAP. The age and gender distribution of all subjects are shown in Fig. 5. The age of COVID-19 subjects is $49 \pm 14$ years, which is significantly younger than that of CAP subjects as $56 \pm 14$ years (p<0.001 with two-sample t-test). There are 856 males and 802 females in COVID-19, which shows slightly higher male to female ratio

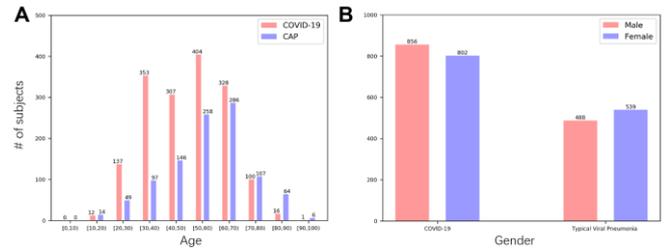

Fig. 5. Distributions of (A) age and (B) gender for the 2685 subjects.

than the CAP with 486 males and 541 females (p=0.038 through Chi-square test).

We developed a machine learning framework to automated process the CT images, extract features, and classification of COVID-19 from CAP patients. The proposed method, named infection size-aware random forest (iSARF), includes a 3-level decision tree to separate subjects into different groups based on the size of infected lesions, and followed by random forests for classification in each group. The performance of the proposed method was evaluated through a 5-fold cross-validation. Comparison methods include logistic regression (LR), support vector machine (SVM), and neural network (NN). The mean ROC was averaged from the 5 folds. Results are shown in Fig. 6. In particular, the proposed method demonstrates an overall superior ROC curve, with the highest AUC of 0.942. Fig. 6B shows the details of metrics, and the proposed method obtains the best performance with sensitivity of 0.907, specificity of 0.833, and accuracy of 0.879. The LR and NN methods perform similarly, and SVM has relatively lower performance.

We further broke down the whole dataset into 5 size groups to better evaluate the performance of methods. Results reveal that, although overall accuracy seems high in the size group of <0.01%, all methods actually have low sensitivity, which means the COVID-19 patients with small infections (most likely in early stage) are quite difficult to be detected. Similarly, all methods show low specificity in the size group of >10%, which suggests the subjects with large infections would be mainly treated as COVID-19. This suggests that, the studies that report only overall performance such as Fig. 6B would easily be biased due to the imbalanced distribution of infected lesion sizes. For the remaining groups, the overall screening performance increases gradually as the infected lesion size becomes larger. For example, accuracy increases from 0.753, 0.819 to 0.907 in the groups of 0.01-0.1%, 0.1-1%, and 1-10%. The individual classification results of the proposed method are also shown in Fig. 7. It could be observed that the sensitivity to screen COVID-19 still have a relatively low performance and needs to be improved especially in the small size groups, e.g., <0.01% and 0.01-0.1%.

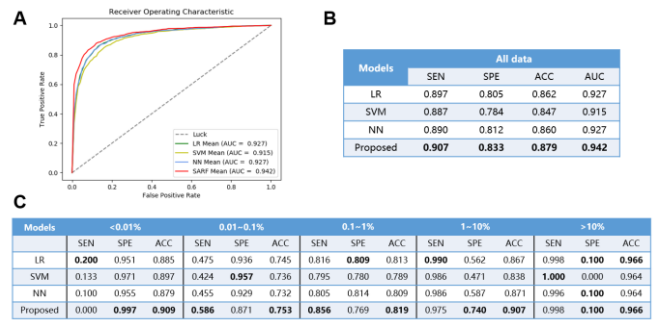

Fig. 6. Illustration of (A) the mean ROC curves from 5 folds in different classifiers, (B) overall performance, and (C) performance after dividing into 5 groups. SEN means sensitivity, SPE means specificity, ACC means accuracy, and AUC means area under curve.

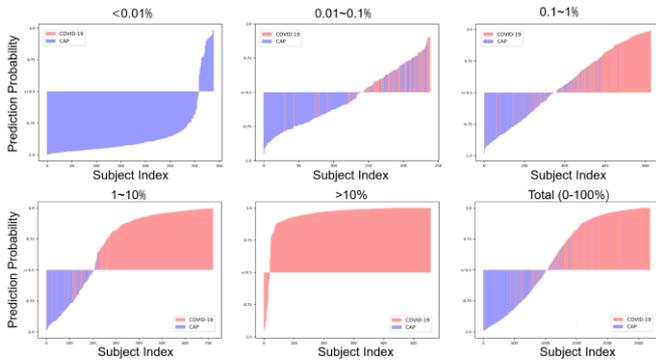

Fig. 7. Illustration of predicted probability with classification labels for different size groups by using the proposed method.

The features picked mostly by the proposed method were shown in Fig. 8. Note that they are weighted by the times of being picked in 5 folds. Results show that the volume and number features are mostly from the lung segments, and several features are from the lung lobes and overall asymmetry. Histogram features are largely selected including many bins, and surface features focus on the short range as 3 and 6 voxels.

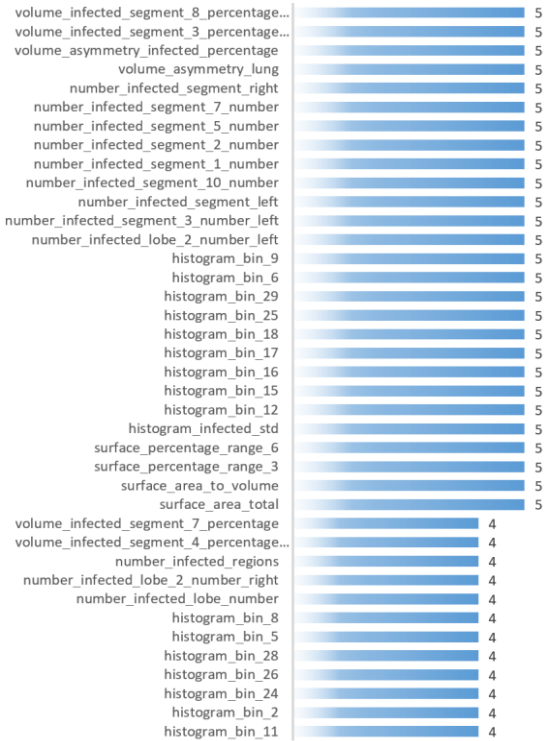

Fig. 8. Illustration of features mostly picked in the proposed model.

## IV. DISCUSSION

Positive results of nucleic acid-based laboratory testing are still the diagnosis standard of COVID-19. Nevertheless, laboratory testing was also reported time-consuming and having a high false-negative rate, and thus CT imaging is suggested as a complementary tool or even recommended approach to the COVID-19 screening in China. A recent study compared RT-PCR results with chest CT scans in 1014 subjects, and found that chest CT scans achieved high sensitivity of 97%, low specificity of 25%, and accuracy of 68%. However, the performance could potentially be improved if the chest CT reading could leverage advanced image analysis techniques. In this study, we proposed a machine learning method with COVID-19 specific features to explore the potential application of CT-based COVID-19 screening. Results are promising, showing 90.7% sensitivity, 83.3% specificity, and 87.9% accuracy. These findings suggest that the proposed method could effectively assist the screening of COVID-19 from CAP.

To date, there is few reported study on machine learning based COVID-19 CT image analysis yet. Wang *et al.* reported a disease screening framework using deep learning method with sensitivity of 74% and specificity of 67%, in a total of 44 COVID-19 patients and 55 typical viral pneumonia patients[12]. Xu *et al.* studied the early screening of COVID-19 from Influenza-A viral pneumonia and healthy cases with totally 618 CT samples[13]. Li et al. performed a study on screening among 468 COVID-19, 1511 CAP, and 1303 non-pneumonia patients. Shi *et al.* presented machine learning model to classify severe patients from non-severe group, where 45 severe cases and 151 non-severe were involved[21].

One of compelling findings of the current study is the unbalanced distribution of infection size between COVID-19 and CAP patients. In accordance with the previous reports[8, 19, 22], the observation regarding the high prevalence of bilateral and peripheral lung opacities, multiple lobular and subsegmental areas infections, and multifocal involvement is the most typical findings on CT images of COVID-19 patients. For example, COVID-19 pneumonia tends to manifest on CT images as bilateral and a slight predominance in the right lower lobes with the mean involved lung segment number of 10.5[3]. Besides, multifocal involvement was more common than unifocal involvement in COVID-19 patients[8]. In the present study, we proposed a size-aware and location-specific machine learning framework. The proposed method first separated the cohorts into different size groups through the decision tree, thus reducing the bias of infection size. From 5-fold experiments, the obtained decision trees show that, when separating the dataset into 4 groups, consistent thresholds around 0.01%, 0.3%, and 7% could be obtained (supplementary material Fig. S1). The feature weight analysis further shows that few size related features were chosen by the proposed method. Therefore, our findings emphasize the importance of percentage of infected lesion in the prediction of COVID-19 pneumonia. Meanwhile, we also compared the current handcrafted features in this study with Radiomics features extracted directly from infected lesions. Results show that, using Radiomics features alone, the proposed method has lower performance (2.4% in accuracy) than that of handcrafted features. By combing the Radiomics with handcrafted features, the performance of the proposed method slightly improved (0.6% in accuracy) (supplementary material Fig. S2).

A few limitations still exist in this study. First, no pneumonia-related clinical characteristics was available to examine the correlation between symptoms and radiologic findings. Including more clinical characteristics might advance our ability to better describe our radiologic findings of COVID-19 pneumonia. Second, only baseline CT findings of COVID-19 patients that clinicians and radiologists first encountered were included, rather than findings from follow-up CT scans. Therefore, a follow-up study of the disease progression is needed in future work. Third, the symptom severity of COVID-19 and differential diagnosis of pneumonia subtypes were not included in this study, which warrants further investigation.

It is worth noting that the proposed method has been integrated into uCloud platform as an online service and available to over 20 clinical facilities in China to date. Meanwhile, our research portals that include computer hardware such as GPUs and software have been deployed and used in over 50 hospitals, in which over 20 are in Wuhan, China.

In conclusion, CT imaging demonstrates high accuracy through machine learning technique and thus could be an efficient tool for COVID-19 screening. The infection size bias needs to be considered in both the method development and the result evaluation process.

REFERENCES

[1] J. T. Wu, K. Leung, and G. M. Leung, "Nowcasting and forecasting the potential domestic and international spread of the 2019-nCoV outbreak originating in Wuhan, China: A modelling study," *Lancet,* vol. 395, no. 10225, pp. 689-697, 2020.
[2] C. Wang, P. W. Horby, F. G. Hayden, and G. F. Gao, "A novel coronavirus outbreak of global health concern," *Lancet,* vol. 395, no. 10223, pp. 470-473, 2020.
[3] H. Shi *et al.*, "Radiological findings from 81 patients with COVID-19 pneumonia in Wuhan, China: A descriptive study," *The Lancet Infectious Diseases,* p. in press, 2020.
[4] E. de Wit, N. van Doremalen, D. Falzarano, and V. J. Munster, "SARS and MERS: recent insights into emerging coronaviruses," *Nature Reviews Microbiology,* vol. 14, no. 8, p. 523, 2016.
[5] A. R. Sahin *et al.*, "2019 novel coronavirus (COVID-19) outbreak: A review of the current literature," *EJMO,* vol. 4, no. 1, pp. 1-7, 2020.
[6] D. Wang *et al.*, "Clinical characteristics of 138 hospitalized patients with 2019 novel coronavirus–infected pneumonia in Wuhan, China," *Jama,* p. in press, 2020.
[7] J. F. Chan *et al.*, "A familial cluster of pneumonia associated with the 2019 novel coronavirus indicating person-to-person transmission: a study of a family cluster," *Lancet,* vol. 395, no. 10223, pp. 514-523, Feb 15 2020.
[8] Y. Li and L. Xia, "Coronavirus disease 2019 (COVID-19): Role of chest CT in diagnosis and management," *AJR Am J Roentgenol,* pp. 1-7, Mar 4 2020.
[9] Y. Fang *et al.*, "Sensitivity of Chest CT for COVID-19: Comparison to RT-PCR," *Radiology,* p. 200432, Feb 19 2020.
[10] K. Wong *et al.*, "Thin-section CT of severe acute respiratory syndrome: Evaluation of 73 patients exposed to or with the disease," *Radiology,* vol. 228, no. 2, pp. 395-400, 2003.
[11] Z. Y. Zu *et al.*, "Coronavirus disease 2019 (COVID-19): A perspective from China," *Radiology,* p. 200490, Feb 21 2020.
[12] S. Wang *et al.*, "A deep learning algorithm using CT images to screen for corona virus disease (COVID-19)," *medRxiv,* 2020.
[13] X. Xu *et al.*, "Deep learning system to screen coronavirus disease 2019 pneumonia," *arXiv preprint arXiv:2002.09334,* 2020.
[14] F. Shan *et al.*, "Lung infection quantification of COVID-19 in CT images with deep learning," *arXiv preprint arXiv:2003.04655,* 2020.
[15] F. Milletari, N. Navab, and S.-A. Ahmadi, "V-net: Fully convolutional neural networks for volumetric medical image segmentation," in *2016 Fourth International Conference on 3D Vision (3DV)*, 2016, pp. 565-571: IEEE.
[16] M. Chung *et al.*, "CT imaging features of 2019 novel coronavirus (2019-nCoV)," *Radiology,* p. 200230, 2020.
[17] D. M. Hansell, A. A. Bankier, H. MacMahon, T. C. McLoud, N. L. Muller, and J. Remy, "Fleischner Society: glossary of terms for thoracic imaging," *Radiology,* vol. 246, no. 3, pp. 697-722, 2008.
[18] X. Li, X. Zeng, B. Liu, and Y. Yu, "COVID-19 infection presenting with CT halo sign," *Radiology: Cardiothoracic Imaging,* vol. 2, no. 1, p. e200026, 2020.
[19] A. Bernheim *et al.*, "Chest CT findings in coronavirus disease-19 (COVID-19): Relationship to duration of infection," *Radiology,* p. 200463, 2020.
[20] F. Song *et al.*, "Emerging 2019 novel coronavirus (2019-nCoV) pneumonia," *Radiology,* p. 200274, 2020.
[21] W. Shi *et al.*, "Deep learning-based quantitative computed tomography model in predicting the severity of COVID-19: A retrospective study in 196 patients," p. in press, 2020.
[22] C. Huang *et al.*, "Clinical features of patients infected with 2019 novel coronavirus in Wuhan, China," *Lancet,* vol. 395, no. 10223, pp. 497-506, Feb 15 2020.

# Supplementary Material for "Large-Scale Screening of COVID-19 from Community Acquired Pneumonia using Infection Size-Aware Classification"

**Results of decision trees in the proposed method**

In each fold, one 3-level decision tree was obtained (Fig. S1). The size of infected lesion was the only feature to automatically decide the split threshold in each node to separate the dataset into two groups. For example, in fold 1, a size threshold of 0.27% was chosen to separate all the training data in this fold into two groups. Then, these smaller size group was again separated by a size threshold of 0.01%. The larger size group was separated by a size threshold of 7.4%. Overall, the thresholds in all 5 folds are in the similar range.

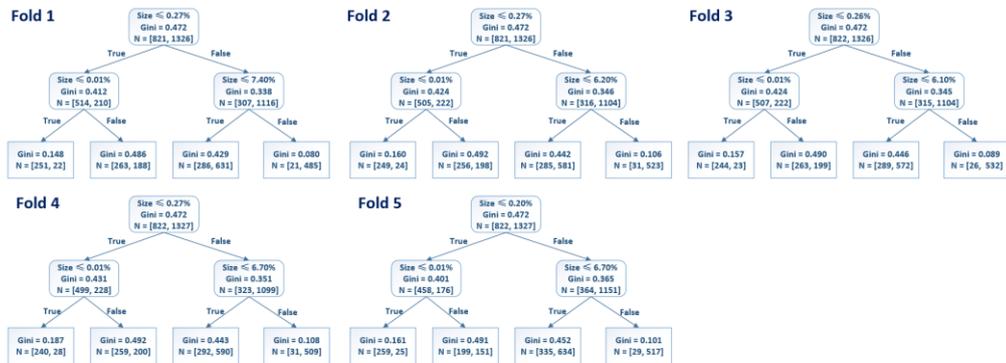

Fig. S1. Results of decision tree of the proposed method in 5 folds.

**Evaluation of the added value of Radiomics features**

A set of 93 Radiomics features were automatically extracted for each subject from the infected lesion from the intensity image. The features include 19 first-order intensity statistics (e.g., average gray level intensity, range of gray values), and 74 texture features (e.g., gray level co-occurrence matrix, gray-level run-length matrix, gray-level size-zone matrix, and neighborhood gray-tone difference matrix).

We compared the performance of using the proposed method with 1) proposed handcrafted features, 2) Radiomics features only, and 3) combined handcrafted and Radiomics features. Fig. S2A shows that, the method with handcrafted features has 2% higher AUC than that using Radiomics features. By combining them together, the AUC is slightly increased for 0.4%. In Fig. S2B, the method with both features has 0.4% lower sensitivity, 2.1% higher specificity, and 0.6% higher accuracy than the method with handcrafted features. The size group analysis found that the combined method shows around 2% accuracy improvement in the 0.01~0.1% group and the 1~10% group.

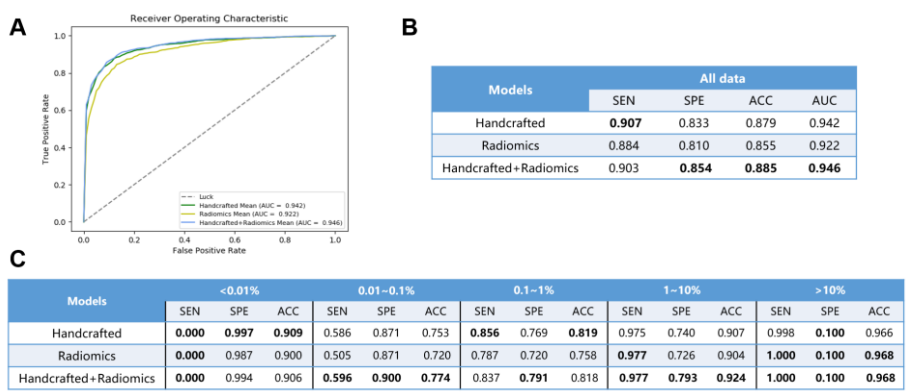

Fig. S2. Illustration of (A) the mean ROC curves from 5 folds in different feature sources, (B) overall performance, and (C) performance after dividing into 5 size groups. SEN means sensitivity, SPE means specificity, ACC means accuracy, and AUC means area under curve.